\documentclass{article}

\pdfoutput=1
\usepackage{arxiv}

\usepackage[utf8]{inputenc} % allow utf-8 input
\usepackage[T1]{fontenc}    % use 8-bit T1 fonts
\usepackage{hyperref}       % hyperlinks
\usepackage{url}            % simple URL typesetting
\usepackage{booktabs}       % professional-quality tables
\usepackage{amsfonts}       % blackboard math symbols
\usepackage{nicefrac}       % compact symbols for 1/2, etc.
\usepackage{microtype}      % microtypography
\usepackage{lipsum}
\usepackage{graphicx}
\usepackage[sort]{cite}

\title{Optimal surface topography for cell adhesion is driven by cell membrane mechanics}

\author{Matej Daniel\\
  Faculty of Mechanical Engineering, Czech Technical University in
  Prague, Czechia\\
  \texttt{matej.daniel@cvut.cz}\\
  \And
  Kristin Eleršič Flipič\\
  Faculty of Mechanical Engineering, Czech Technical University in
  Prague, Czechia\\
  \And
  Eva Filová\\Institute of Experimental Medicine of the Czech Academy of Sciences, Czechia\\
  \And
  Jaroslav Fojt\\
   University of Chemistry and Technology, Prague, Czechia,
}
\begin{document}
\maketitle

\begin{abstract}
  Titanium surface treated with titanium oxide nanotubes was used in many
  studies to quantify the effect of surface topography on cell fate. However,
  the predicted optimal diameter of nanotubes considerably differs among
  studies. We propose a model that explain cell adhesion to nanostructured
  surface by considering deformation energy of cell protrusions into titanium
  nanotubes and adhesion to surface. The optimal surface topology is defined as
  a geometry that gives membrane a minimum energy shape. A dimensionless
  parameter, the cell interaction index, was proposed to describe interplay
  between the cell membrane bending, intrinsic curvature and strength of cell
  adhesion.
  Model simulation show that optimal nanotube diameter ranging from 20 nm to 100
  nm (cell interaction index between 0.2 and 1, respectively) is feasible within
  certain range of parameters describing adhesion and bending energy. The
  results indicates a possibility to tune the topology of nanostructural surface
  in order to enhance proliferation and differentation of cells mechanically compatible
  with given surface geometry while suppress the growth of other mechanically
  incompatible cells.

\end{abstract}

% keywords can be removed
\section{Introduction}

Strong bonds between the implant and bone cells \cite{Feller2015,Imani2012} is
required for the long-term stability of the implant in the human body
\cite{Souza2019}. It was shown that bone cellular response is directly affected
by titanium surface characteristics like roughness, chemistry, wettability or
more recently studied surface topography~\cite{Shokuhfar2014}. Various methods
for surface modification were employed in order to promote cell–substrate
interactions~\cite{Martinez-Marquez2020, Minagar}.

\begin{figure}[!ht]
  \centering
  \includegraphics[width=.8\textwidth]{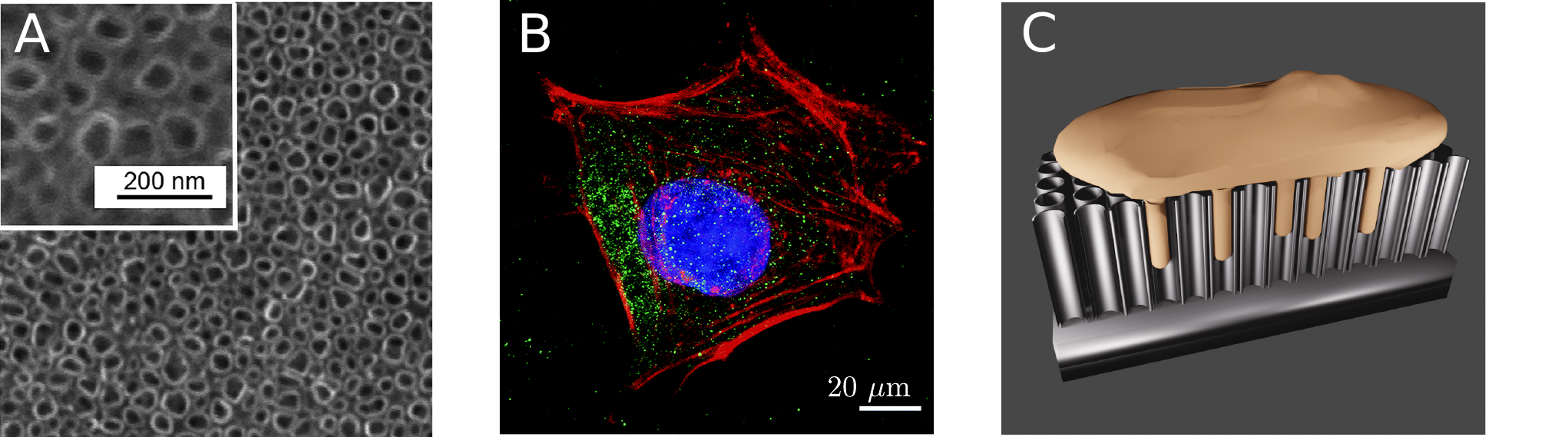} 
  \caption{(A) Titanium nanotubes on cpTi of average diameter 66 $\pm$ 17 nm and
    length 1097 $\pm$ 75 nm. (B) Immunofluorescence staining of talin in human
    Saos-2 osteoblast-like cells on nanostructured surface. (C) Schematic view
    of cell anochored into nanostructured surface.  A,B adopted from
    Voltrova et al, 2019. \cite{Voltrova2019}}
  \label{fig:talin}
\end{figure}

The anodic oxidation is adopted to create a nanostructured titanium
surface~\cite{Li2018} by formation of TiO$_2$ nanotubular structures
(TNTs)~\cite{Souza2019, Frandsen2013} (Fig.~\ref{fig:talin}A). TNTs increase 
surface area that favors bone deposition and could improve therapeutic
efficiency by serving as a reservoir for drug delivery~\cite{Hao2013,Wei2019}.
The advantage of anodic oxidation is that the diameter, the wall thickness and
the length of TNTs can be controlled by the process variables such as electrical
current power, anodization time, temperature, applied potential, and electrolyte
chemical composition~\cite{Li2018, Hamlekhan2014a, Indira2015}. TNTs length can
range from 0.1 up to 1000$\mu$m while the inner diameter can range from 7 to 150
nm~\cite{Souza2019, Paulose2007,Peng2019}.

Surface treated with TNT array present a controlled environment that allows to
quantify the effect of surface topography on cell fate~\cite{Kulkarni2015a}.
Nanotube diameter, rather than the other characteristics of the surface layer,
exhibits critical impact on cell adhesion and proliferation~\cite{Park2007,
  Frandsen2013, Hao2013, Minagar2013, Tian2015}. It was further suggested that
there exists an optimal diameter for TNTs that enhance osteointegration
\cite{Park2009}. However, estimated values of the optimal diameter are
contradictory. Park et al., 2017 \cite{Park2007} reported the optimal nanotube
diameter to be 15 nm based on mesenchymal stem cell proliferation on TNT
surface. They also report that the cell adhesion and spreading decreases on TNT
layers with a tube diameter larger than 50 nm. Yu et al., 2010 \cite{Yu2010}
found that MC3T3-E1 preosteoblast adheres well on TNTs of diameter 20–70 nm
while the cell attachment is low on TNTs of diameter 100-120 nm. Similar
behavior was observed for oestoblast-like MG-63 cells that exhibit higher
spreading on 30 nm TNTs whereas the larger diameter of 90 nm had the worst cell
viability\cite{Hao2013}. Both glioma and osteosarcoma cells exhibit optimal cell
adhesion, migration, and proliferation on 20 nm TNTs \cite{Tian2015}. Limited
spreading on larger diameter TNTs was also reported for malignant cancer cells
(T24) of urothelial origin \cite{Imani2012}. Osteogenic differentiation of
primary rat osteoblasts was observed on 35 nm (amorphous phase) and 41 nm
(anatase phase) surface \cite{Khrunyk2020}. Das et al., 2009 \cite{Das2009}
found 2–3 fold increase in human osteoblast attachment and spreading on 50
nm-diameter TNTs surfaces in comparison to flat Ti samples.  Oh et al. reported
 improved adhesion of hMSC on 30 nm TiO$_2$ nanotubes and improved osteogenic
 differentiation on nanotubes with a diameter of 70 and 100 nm \cite{Oh2009, Oh2008}.
MC3T3-E1
osteoblast cells accelerates in the growth on 70 nm TNTs \cite{Li2006}. Brammer
et al, 2009 \cite{Brammer2009} proposed that bone-forming ability of osteoblasts
is higher if grown on TNTs of 100 nm diameter. Also Filova et al, 2015
\cite{Filova2015} and Voltrova et al, 2019 \cite{Voltrova2019} concludes that
optimal diameter of TNTs is around 70 nm for Saos-2 osteoblast-like cell
(Fig.~\ref{fig:talin}B).

 The divergence in results
could be either caused by variations in surface topography and chemistry, due to
individual fabrication protocols or by methods to assess cellular activities
\cite{Park2009a}. It is also likely, that the type of cell line affect optimal
TNT's diameter~\cite{Park2009}. While the preference of cells to small diameter
TNTs (up to 30 nm) could be explained by integrins packing \cite{Park2007, Gongadze2013},
mechanism of adherence to large diameter has not been explained yet.  
% The difference between the studies indicate an inappropriate understanding of
% the cell-TNT interactions~\cite{Martinez-Marquez2020}.
It was suggested that migration of the cell membrane inside the crystalline
nanotubes could be crucial for strong attachment~\cite{Shokuhfar2014,
  GuadarramaBello2017, Souza2019}. The cell protrusions into nanotubes could
strengthen the adhesive interaction of cells with the surface, and thereby
potentially trigger cellular cascades that regulate cell behavior and
differentiation \cite{GuadarramaBello2017}. Cell protrusion into TNTs increases
contact area for attachment but requires extensive membrane deformation into
tubular like structure (Fig.~\ref{fig:talin}C). The aim of the present study is
to quantify the overall energy cost of formation of cell protrusion into TNT.
The hypothesis based on the experimental results is, that there exist an optimal
diameter given by minimum of membrane protrusion energy.

\section{Methods}

\begin{figure}[!ht]
  \centering
  \includegraphics[width=.8\textwidth]{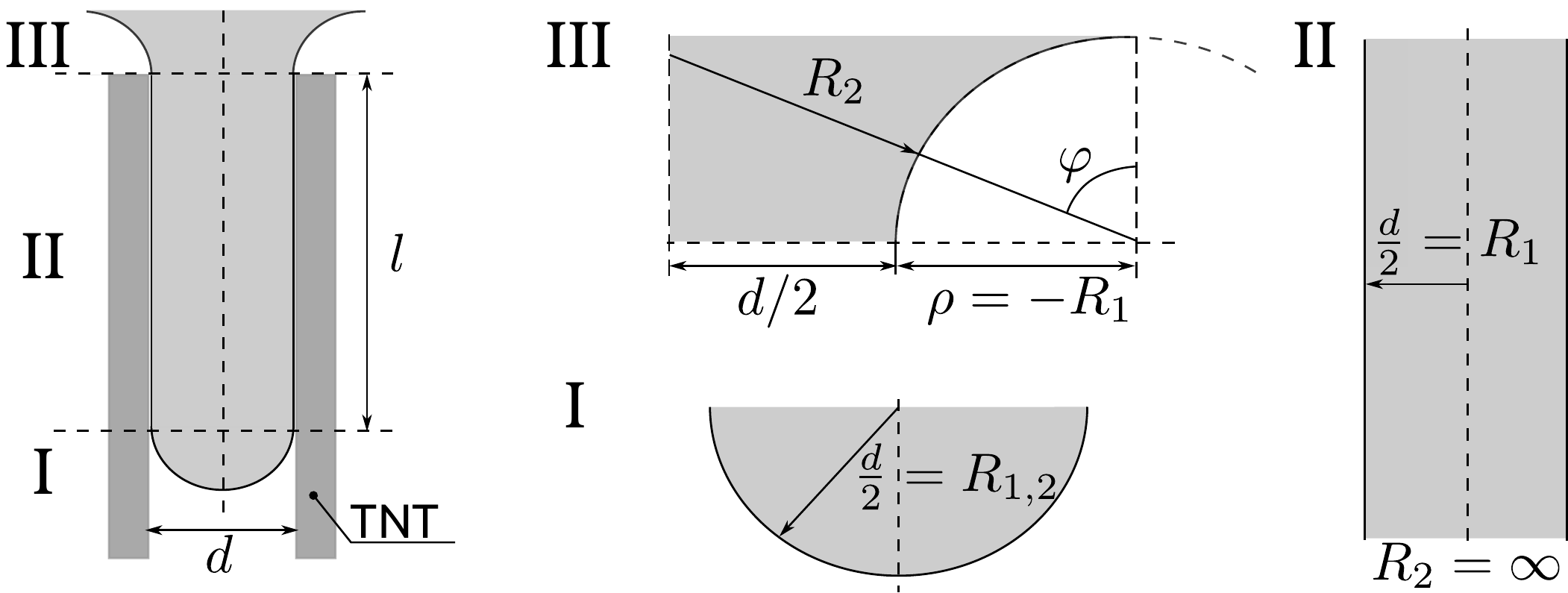}
  \caption{Parametrization of membrane protrusion into TiO$_2$ nanotubule of
    diameter $d$. The shape is divided into three parts: (I) the spherical cup
    of radius $d/2$, (II) the cylindrical segment of length $l$ and radius $d/2$
    and (III) the axisymmetrical collar. Principal curvature $R_1$ and $R_2$ are
    depicted in individual segments.}
  \label{fig:geometry}
\end{figure}

The membrane protrusion into hollow nanotubular structure is assumed to be
axisymmetric and its dimensions are determined by the shape of the nanotube. The
membrane therefore forms a hollow cylinder of diameter $d$ closed by a
hemispherical cup and joined to the central body along the contour shown in
Fig.~\ref{fig:geometry}. Two energy contribution are considered in mechanics of
nanotubular protrusion: the adhesion energy $F_a$ between the TNT inner surface
and membrane and the deformation energy $F_b$ of the cell
membrane~\cite{Sackmann2014}. % The free energy is expressed as a sum of these
% energies.
% \begin{equation}
%   \label{eq:free_energy}
%   F = F_a + F_b
% \end{equation}

% The interaction between the cell membrane and the TNT inner surface can be
% expressed by the adhesion energy $F_a$.
The adhesion energy is defined as the
excess energy released after the cell attaches to the surface~\cite{Feller2015}.
Surface energy quantifies the formation of intermolecular bonds and depends on
the contact area $A$ with a proportionality constant $\gamma$. According to our
model (Fig.~\ref{fig:geometry}), the cell membrane is in contact with the TNT
only in its central tubular segment (II). The adhesion energy could therefore be
expressed as
\begin{equation}
  \label{eq:adhesion_energy}
  F_a =  - \gamma \pi d l
\end{equation}
where the minus sign denotes energy release after adhesion.

Formation of protrusion requires deformation of the membrane from the mostly
planar shape into the shape of thin cylinder. The bending energy of the membrane
is commonly described by Helfrich energy \cite{Helfrich1973}. The elastic strain
energy proposed by Helfrich depends on the mean ($H$) and Gaussian curvature. As
we do not expect the change in cell topology by protrusion formation, the Gauss
term could be neglected because of Gauss-Bonnet theorem~\cite{Bassereau2018}.
\begin{equation}
  \label{eq:Helfrich}
  F_b = \frac{1}{2} k_b \int\limits_A (2\,H - C_0) dA
\end{equation}
where $k_b$ is the bending modulus of cell membrane and $C_0$ is the spontaneous
curvature. Spontaneous curvature, or more precisely the spontaneous mean
curvature, present a penalty for the mean curvature asymmetry
\cite{Chabanon2018}. The mean curvature can be expressed as an average of
principal curvature values $C_1$ and $C_2$ defined as the inverse values of
corresponding radii of curvatures $R_1$ and $R_2$, respectively (Appendix~\ref{sec:appendix}). 

In order to get insight into the interaction between bending and adhesion, we
will analyze equilibrium of part II in Fig.~\ref{fig:geometry}.
Contribution of part I and III could be neglected if the length of the cylinder
$l$ is much greater than the diameter, i.e. $l \gg d$. The free energy is
expressed from Eqs.~(\ref{eq:adhesion_energy}) and (\ref{eq:FbII}).
\begin{equation}
  \label{eq:energywithoutcup}
  F = \frac{1}{2} k_b \pi l \left(\frac{4}{d} - 4\,C_0 + C_0^2\,d  \right)  - l \gamma \pi d 
\end{equation}
The central assumption is, that the membrane attains a shape that minimizes the
overall energy. % The membrane protrusion is energetically favorable if free
% energy in Eq.~(\ref{eq:energywithoutcup}) is lower than zero. The minimal
% diameter of the nanotube that facilitates ingrowth $d_0$ is
% \begin{equation}
%   \label{eq:diameter_ingrowth}
%   \frac{2}{d_0} = C_0 + \sqrt{\frac{2\gamma}{k_b}}
% \end{equation}
% It should be noted, that the diameter $d_0$ does not depends on the nanotube
% length $l$. The real value of $d_0$ differs from
% Eq.~(\ref{eq:diameter_ingrowth}) as we neglected contribution of part I
% (Eq.~(\ref{eq:FbI})) and III (Eq.~(\ref{eq:FbIII})).
We may further assume, that
there exist an optimal diameter that corresponds to energy minimum. The minimum
of the energy present a stationary point and could be expressed using interior
extremum theorem.
\begin{equation}
  \label{eq:stationary_point}
  d_0 = \sqrt{\frac{4\,k_b}{k_b C_0^2 - 2 \gamma}}
\end{equation}
The value of optimal diameter depends on adhesion constant and bending rigidity
of the membrane. The optimal diameter exists if intrinsic curvature is higher
than a threshold value. We denote this value as a critical curvature
$C_{\mathrm{crit}}$.
\begin{equation}
  \label{eq:conditionC0}
  C_{\mathrm{crit}} = \sqrt{\frac{2\,\gamma}{k_b}}
\end{equation}
To describe interaction between the cell protrusions and nanostructured surface,
we define a dimensionless number $I_c$ denoted as cell interaction index.
\begin{equation}
  \label{eq:interactionIndex}
  I_c = \frac{C_\mathrm{crit}}{C_0}
\end{equation}

As shown above, the energy of cell membrane TNT interaction depends on
mechanical properties of membrane described by the bending modulus $k_b$ and the
spontaneous curvature $C_0$ and on interaction between membrane and TNT surface
described by density of surface energy $\gamma$. The bending modulus of the cell
range from 5 $k_BT$ for phospholipid membrane \cite{Morshed2020} to 200 $k_BT$
for cells \cite{Pontec2011}, where $k_B$ is the Boltzmann constant. In the
previous study of osteoblasts mechanics, the value of 100 $k_BT$ was used to
describe osteoblasts bending rigidity~\cite{Gongadze2013}. % The TiO$_2$ surface
% energy density $\gamma$ for water adsorption ranges from 0.7 to 2.2
% J\,m$^{-2}$ for anatase and rutile, respectively.
The cell binding energy per unit area
$\gamma$ may range from 0.05 to 56 mJ\,m$^{-2}$ for various cell
types~\cite{Winklbauer2019}. % The $\gamma$ value of 50~mJ\,m$^{-2}$ was proposed
% to describe adhesion of osteoblasts to the TiO$_2$ surface~\cite{Luo2018}.
The spontaneous curvature of the cell membrane is determined by lipid
composition and interactions between lipids and proteins \cite{McMahon2005}. It
could have either positive values (intrinsic bending inwards) or negative values
(bending outwards). It was reported that lipid bilayer spontaneous curvatures
ranges from -0.2 to 0.2 nm$^{-1}$ \cite{Yesylevskyy2017}.

%%%%%%%%%%%%%%%%%%%%%%%%%%%%%%%%%%%%%%%%%%
\section{Results}

\begin{figure}[!ht]
  \centering
  \includegraphics[width=.5\textwidth]{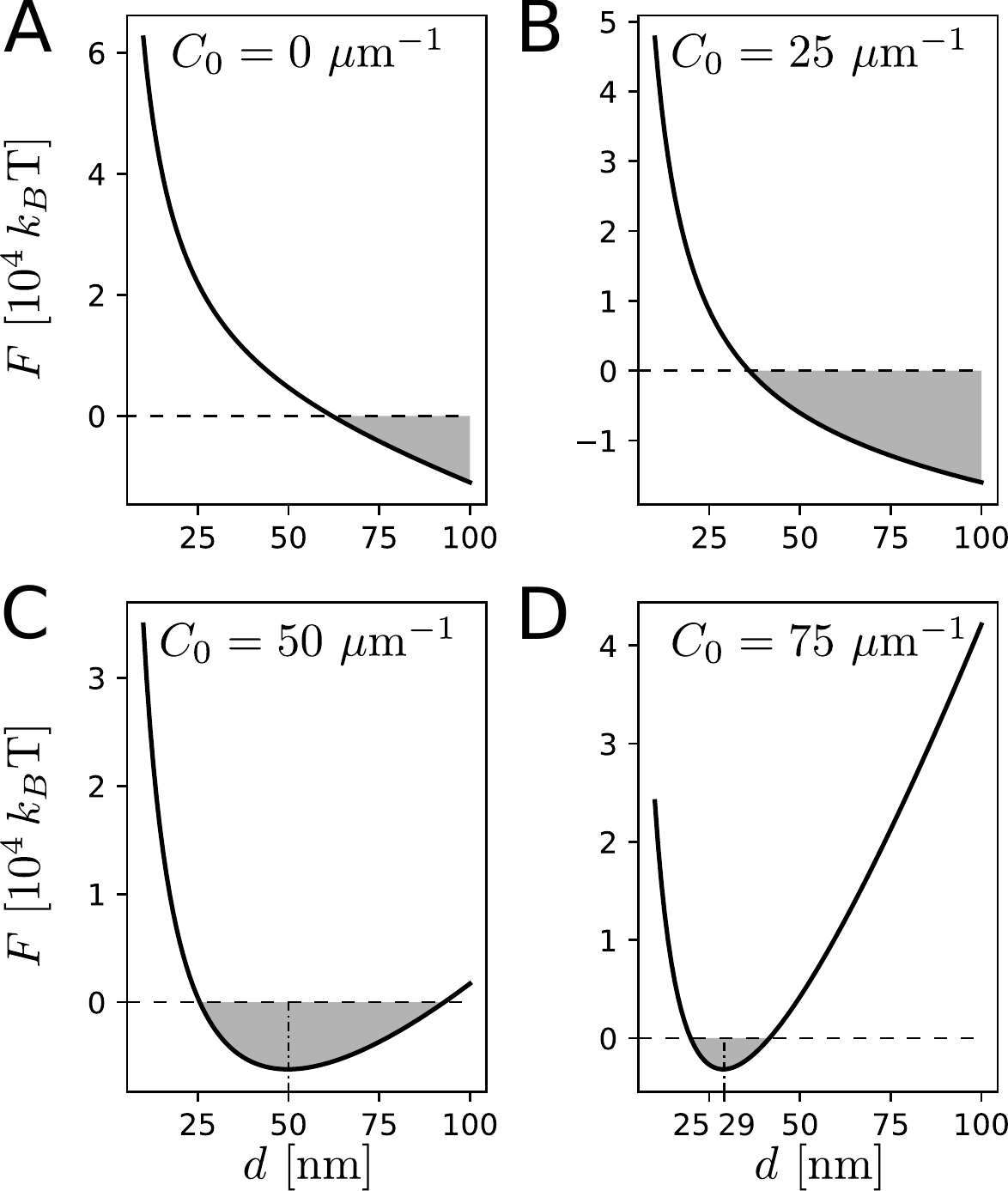}
  \caption{Free energy of membrane protrusion ($F$) into TNT: (A,B) spontaneous
    curvature $C_0$ is lower than the critical curvature $C_\mathrm{crit}$,
    (C,D) spontaneous curvature $C_0$ is higher than the critical curvature
    $C_\mathrm{crit}$. Gray region indicate area where formation of protrusion
    is energetically favorable. The critical curvature $C_\mathrm{crit}=
    35~\mu\mathrm{m}^{-1}$ and energy is calculated for $k_b=100\,k_B$T, $\gamma
    = 0.25 $mJ\,m$^{-2}$, $l=1000$~nm for protrusion shape shown in Fig.~\ref{fig:talin}.}
  \label{fig:1}
\end{figure}

The existence of optimal diameter of cell for attachment into TNTs depends on
the difference between the spontaneous $C_0$ and the critical curvature. If
$C_0$ is lower than $C_{\mathrm{crit}}$, there is no optimal diameter and cells
migrate into TNTs larger than threshold. However, if $C_0$ is higher than
$C_{\mathrm{crit}}$, there exists a limited range of TNTs' diameters in which
the formation of membrane protrusion is energetically convenient (Fig.
\ref{fig:1}C,D). For higher spontaneous curvature, the TNTs' optimal diameter
range is smaller and the energy rises considerably for larger diameters
(Fig.~\ref{fig:1}D).

\begin{figure}[!ht]
  \centering
  \includegraphics[width=.55\textwidth]{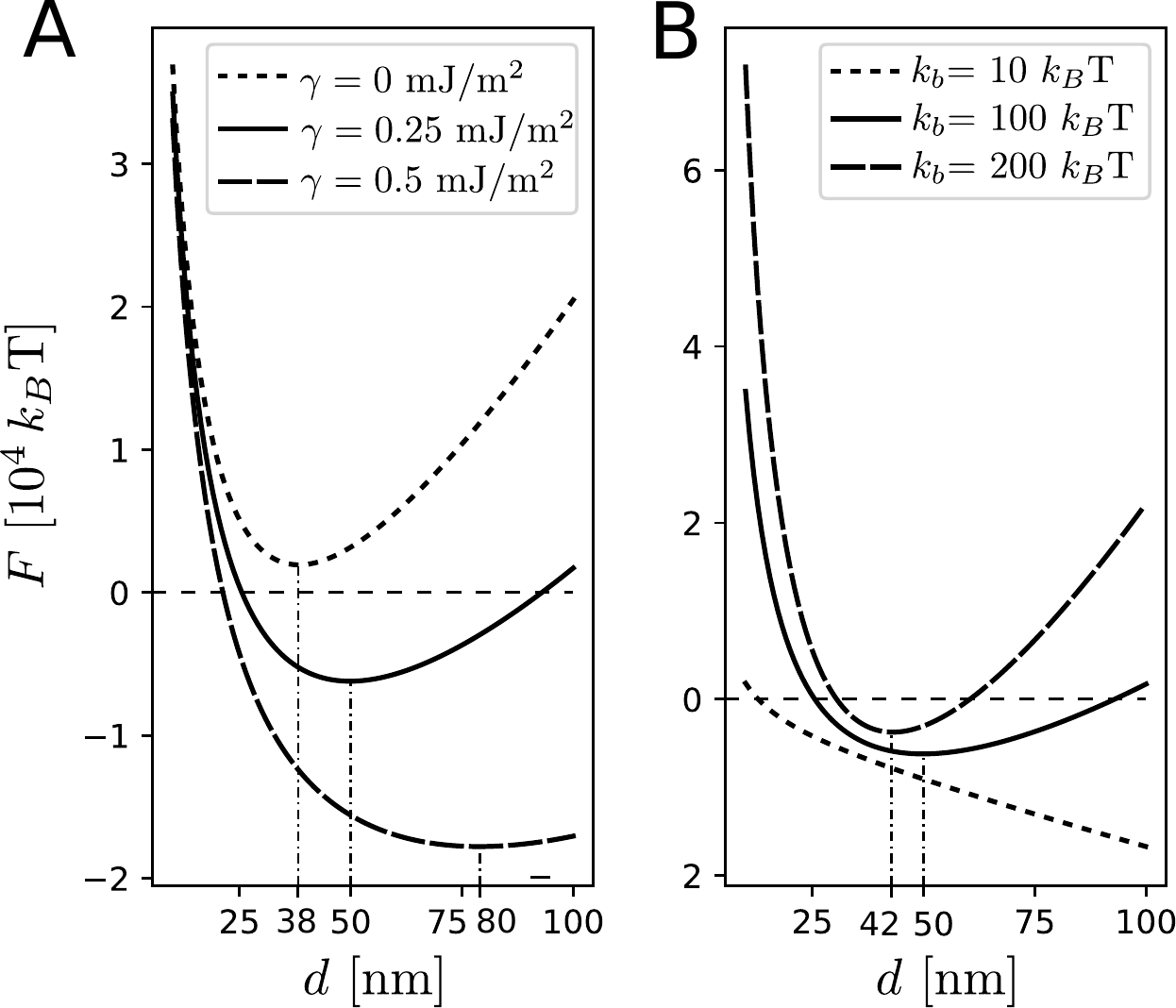}
  \caption{The effect of (A) surface energy density $\gamma$ and (B) bending
    modulus of the membrane $k_b$ on free energy minimum for
    $C_0=50\mu\mathrm{m}^{-1}$ and $l=1000$~nm. Solid line correspond to
    Fig.~\ref{fig:1}C. Optimal diameter is depicted for each curve.}
  \label{fig:2}
\end{figure}

The critical curvature is a function of binding energy per unit area $\gamma$
and bending stiffness of membrane $k_b$ (Eq.~(\ref{eq:conditionC0})). Increase
in adhesion strength (Fig.~\ref{fig:2}A) and decrease in bending stiffness
(Fig.~\ref{fig:2}B) enhance formation of cylindrical protrusion by lowering
membrane free energy. For stiff membrane or limited adhesion between the cell
membrane and the TNTs' wall, the migration of membrane into TNTs' is not likely
to happen spontaneously (Fig.~\ref{fig:2}A,B).

\begin{figure}[!ht]
  \centering
  \includegraphics[width=.5\textwidth]{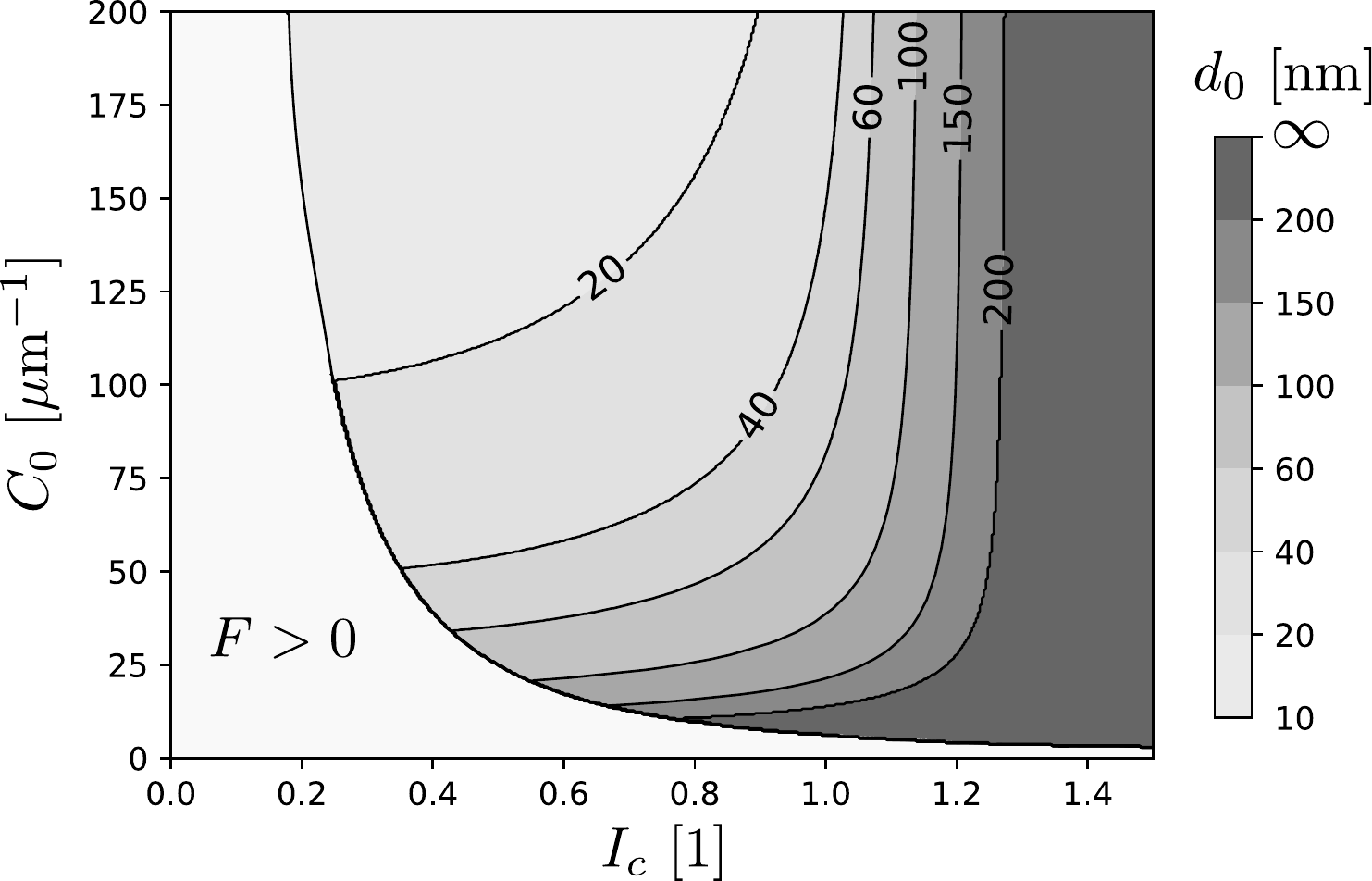}
  \caption{The optimal diameter as a function of interaction index
    $I_c = C_\mathrm{crit}/C_0$.}
  \label{fig:3}
\end{figure}

Figure~\ref{fig:3} shows the dependence of the optimal diameter of TNT ($d_0$)
on the cell interaction index $I_c$, Eq.~(\ref{eq:interactionIndex}). For small
values of $I_c$, the contact between nanostructured surface and cell will not be
formed as the energy required to bend the membrane is higher than the energy
gained in forming adhesion bonds. For $I_c$ between 0.2 and 1, the optimal
topology exists and it depends on the spontanous curvature. Cells with high
spontanous curvature will prefer smaller diameter of TNTs. If $I_c$ is higher
than one, the cell will prefer smooth surface against curved one.

\section{Discussion} 

% Previous experimental studies provide ambiguous values of optimal diameter of
% nanotubular titanium surface for cell growth.
We have hypothesized, that the cell membrane mechanics determines optimal
topology of titanium nanostructured surface. The optimal surface topology is
defined as a geometry that forms membrane into minimum energy shape. Cell
membrane free energy accounts for the cost of the bending energy and the gain in
the adhesion energy. The model explains previous experimental studies
providing ambiguous values of optimal diameter of TNTs for cell growth. Model
simulation show that either small diameters as observed by Park et al, 2007
\cite{Park2007} (Fig.~\ref{fig:1}D) or larger diameters reported by Brammer et
al, 2009 \cite{Brammer2009} (Fig.~\ref{fig:1}C) are feasible within certain
range of parameters describing adhesion and bending
energy. % That optimal diameter $d_0$
% forms a small window, if the $d_0$ is low as described by Park et al, 2009 and
% shown in Fig.~\ref{fig:1}D.

Model analysis indicate that the spontaneous curvature relative to critical
curvature (Eq.~\ref{eq:conditionC0}) determines existence of optimal surface
topology given by optimal diameter $d_0$. However, critical curvature value does
not discriminate whether the cell is or is not attached to the surface.
Therefore, we have defined a new dimensionless parameter describing the
interactions between the cells and the nanostructured surface, the cell
interaction index $I_c$ (Eq.~(\ref{eq:interactionIndex})). The cell interaction
index shows, that a certain parameters range describing cell mechanics
predispose the cell to form stable protrusions into nanostructured surface of
specific topology. For example, it was reported that proliferation of vascular
smooth muscle cells is higher on flat TiO$_2$ surface while the endothelial
cells prefer TNT surface~\cite{Peng2009, Junkar2020}. Experimenal study show
that smooth muscle cells loss their affinity to TNTs after plasma treatment
\cite{Junkar2020}. Plasma treatment is known to increase surface energy and
therefore increase also $C_\mathrm{crit}$ and $I_c$. High $I_c$ corresponds to
minimum energy at flat surface in Fig.~\ref{fig:3} in agreement with experiment.
On the other hand, preference to curved nanostructured surface can be caused
either by high bending modulus, low adhesion or high spontaneous curvature. The
latter was reported to be high in endothelial cells~\cite{Schmid-Schonbein1995}.
The preference to small diameter TNT surface was also observed in cancer cells
\cite{Imani2012, Tian2015} that generally have lower adhesion strength
\cite{Fuhrmann2017} and therefore low $I_c$ in Fig.~\ref{fig:3}.

However, if the adhesion is too low or bending rigidity too high ($I_c$ close to
zero), the cell will not adhere to the surface (Fig.~\ref{fig:3}). For example,
TNT surface decrease the adherence of all bacteria \cite{Ercan2011,
  Puckett2010}. Gram-positive bacteria is surrounded by a bacterial wall of
stiff glycan strands cross-linked into lipid bilayer \cite{Harper2020}. This
composite structure considerably increase bending rigidity~\cite{Auer2017}. High
bending rigidity implies low $I_c$ and limited protrusion into TNT wall
(Fig.~\ref{fig:3}). For bacteria, TNT surface has small contact area restricted
to the terminal ends of nanotubes as protrusion formation is energetically
unfavourable.

Previous studies on TNT bioactivity focus mostly on material properties like
surface chemistry, crystallinity, nanotube size, or water contact angles. The
current study supplement previous research by study the adhesion from the
perspective of the cell while the cell-substrate interactions are described by
the binding energy (Eq.~\ref{eq:adhesion_energy}). As-synthesized TNT are an
extension of the amorphous TiO$_2$ layer~\cite{Martinez-Marquez2020} and after
heat treatment the crystallinity of TiO$_2$ is improved~\cite{Ercan2011}.
Titanium crystallinity (amorphous versus anatase structures) enhances mechanical
strength and increases hydrophilicity, which might improve cell adhesion and
proliferation~\cite{Mazare2012,Khrunyk2020}. Our results indicate, that the high
cell adhesion itself (higher $I_c$) is required for cell attachment, but not
inevitable for having an optimal TNT diameter. The same holds for the water
contact angle that is another measure of surface energy~\cite{Hamlekhan2014a}. 

The model was intentionally kept simple for clarity. However, there are many
other parameters and mechanisms that could be considered in description of
cell-nanosurface interactions. The adherence is described by a single
adhesion energy constant $\gamma$. The cell adhesion is complex process
facilitated by charged protein-mediators~\cite{Gongadze2013}. The adhesion of
proteins is shown to be higher for larger diameter TNT that could farther
facilitate adhesion~\cite{Gongadze2011}. The increase in surface charge could
enhance protein adhesion and promotes osteoblast cell proliferation
\cite{Bandyopadhyay2019}. % The model could be improved by changing considering
% $\gamma$ as a function of above mentioned parameters.
Spontaneous curvature of the membrane $C_0$ is one of the main investigated
parameters within the current study (Figs.~\ref{fig:1}, \ref{fig:3}). While the
spontaneous curvature of lipid bilayer is mostly determined by its lipid
composition, local spontaneous curvature is driven by trans-plasma-membrane or
peripheral proteins~\cite{Yesylevskyy2017}. Therefore, the spontaneous curvature
may not be constant, but it is likely to change along protrusion. In addition,
proteins not only generate curvature, but can also sense membrane curvature
\cite{Breuer2019} and accumulate at curved membrane area~\cite{Strahl2015}.
Similarly, microgrid topography of TiO2 stimulated hMSC adhesion and spreading
area while nanotopography favoured hMSC motility, and osteogenic
differentiation \cite{Chen2018}. 

It is well accepted that the deformation of cell membrane, interacting with the
attached cytoskeleton, affects cell proliferation and
differentiation~\cite{Pontec2011}. For cell adhesion, complex network of
transmembrane integrins and cytoplasmic proteins is of the utmost
importance~\cite{Luo2018}. Extracellular components of integrins attach to
extracellular matrix while their intracellular components are attached to
F-actin through adapter proteins~\cite{Kanchanawong2010} and may directly affect
cell nucleus shape \cite{Anselme2018}. Park et al, 2007 \cite{Park2007} proposed
a hypothesis, that optimal diameter of nanotubes is determined by integrin size.
The size of extracellular domain of integrins is about 10 to 12 nm
\cite{Kanchanawong2010} and thus close integrins packing results in optimal
integrin activation. This hypothesis is supported by the measurements showing
that the 15-20 nm spacing is optimal for cell adhesion, proliferation,
migration, and differentiation \cite{Park2007, Park2009}. This theory was
further implemented into the mathematical model of osteblast
adhesion~\cite{Gongadze2013}. The model well explains narrow window of optimal
diameter observed by Park et al, 2009 \cite{Park2009} but
cannot explain stability of larger diameters \cite{Frandsen2013}.

\begin{figure}[!ht]
  \centering
  \includegraphics[width=.4\textwidth]{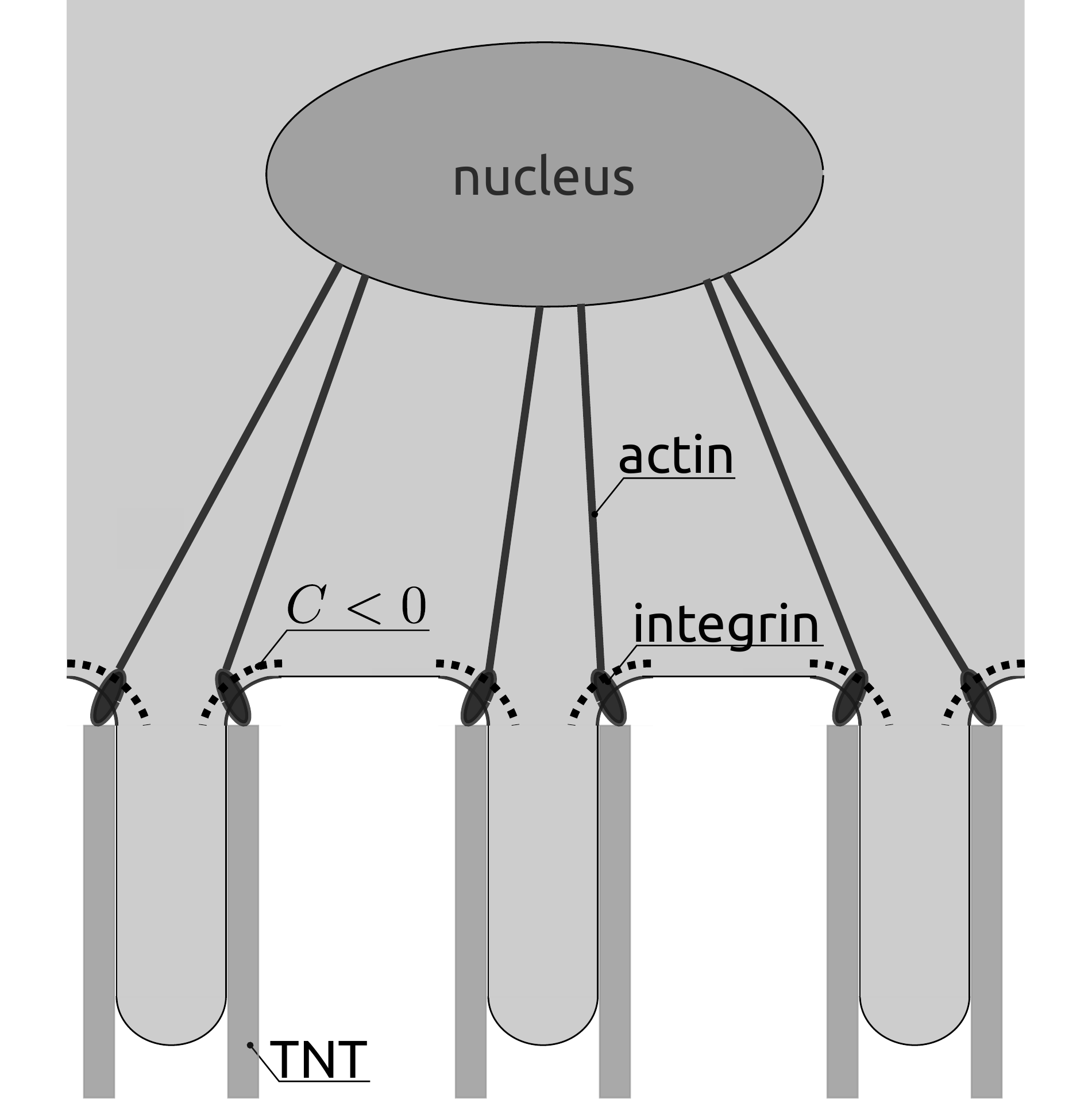}
  \caption{Mechanism of integrin activation at negative curvature area caused by
  cell membrane migration into TNT. Actin transmit the information on focal adhesion
  to nucleus.}
  \label{fig:theory}
\end{figure}

The integrins are also sensitive to the membrane mechanical state including the
curvature~\cite{Kechagia2019}. It was shown, that higher concentration of
integrins occurs at the neck of protrusive podosome-like structures if the
substrate is porous~\cite{Baranov2014}. Podosome neck correspond to part III in
Fig.~\ref{fig:geometry}. It is reasonable to assume, that the same shape of
membrane within TNT will provide similar accumulation of integrins. It was
proposed, that negative membrane curvature increases separation of integrin
cytoplasmic tails, which is known to promote integrin
activation~\cite{Meshik2019}. Therefore we complement a hypothesis of Park et
al, 2007 \cite{Park2007} by adding the role of membrane protrusions into TNTs.
The nanostructured protrusion induce negative curvature in the neck (part III in
Fig.~\ref{fig:geometry}). Area of negative curvature results in accumulation of
integrins and their activation. Actin filaments transmit the focus adhesion
signal to the nucleus activating nuclear mechanotransduction
pathways~\cite{Anselme2018}. Park et al, 2017 observed no focal contact
formation for larger TNT diameters. According to Fig.~\ref{fig:1}D, no
protrusion is formed ($F>0$) and therefore no region of negative curvature
enhancing focal contact exists. This theory is in agreement with molecular
dynamics simulation showing that nanopore-induced membrane curvature increases
bioactivity locally at the neck region~\cite{Belessiotis-Richards2019}.

\section{Conclusions}

The formation of membrane protrusions into TiO$_2$ nanotubes was assessed by
means of cell membrane free energy. Dimensionless parameter, the cell
interaction index $I_c$, was introduced to describe interplay between the cell
membrane mechanics and the nanostructured surface topology. If $I_c$ is close to
zero, no membrane protrusions are formed and no cell adhesion occurs. For $I_c$
greater than one, the cells prefer flat surface. For $I_c$ approximately between
zero and one, there exist an optimal diameter of TNT for given cell line. This
study provides a theoretical basis explaining ambiguous results of experimental
studies reporting wide range of suitable TNT diameters. It was proposed, that
negative curvature region at the neck of membrane protrusion may result in
integrin activation and subsequent cell proliferation. The results
indicates a possibility to tune the topology of nanostructural material in a way to
enhance proliferation and differentation of one cell type that is mechanically
compatible with given surface geometry while suppress the growth of other
mechanically incompatible cells.

\section*{Funding}
\label{sec:fun}

This research was funded by Czech grant agency grant number 16-14758S.

\bibliographystyle{unsrt}
\bibliography{/home/madan/Documents/Bibliography/Bibtex/library.bib}

\newpage

\appendix
\section{Bending energy of membrane protrusion}

\label{sec:appendix}
Energy of membrane forming tubular structure depends on curvature of individual
parts depicted in Fig.~\ref{fig:geometry}. According to the curvature, the
membrane protrusion could be divided into three parts (Fig.~\ref{fig:geometry}).
The first segment correspond to the hemispherical cup, where both principal
curvatures equals to $2/d$ and the energy of the first segment could be
expressed as
\begin{equation}
  \label{eq:FbI}
  F_{bI} = k_b \pi \left(4 - 2\,d\,C_0 + \left( \frac{d \,C_0}{2}  \right)^2 \right)
\end{equation}
The free energy of the central cylindrical part (Fig.~\ref{fig:geometry}, II) is
determined by its length $l$ while the first and the second membrane curvature
are $2/d$ and $0$, respectively.
\begin{equation}
  \label{eq:FbII}
  F_{bII} = \frac{1}{2} k_b \pi l \left(\frac{4}{d} - 4\,C_0 + C_0^2\,d  \right)
\end{equation}
The last part presents a neck, that connect a protrusion to the cell. The neck
is modeled as axisymmetrical structure with one radius of curvature equal to
$\rho$ (Fig.~\ref{fig:geometry}, III). The first curvature is negative as the
membrane bends outwards, $C_1= - 1/\rho$. The second radius of curvature depends
on the distance from axis of symmetry and could be expressed as
$C_2 = \sin(\varphi) / (d/2 + \rho\,(1-sin(\varphi)))$ \cite{Boal2002} where
$\varphi$ is defined in Fig.~\ref{fig:geometry}. For the sake of simplicity, we
further assume that the radius $\rho$ equals to $d/2$. The energy could be
expressed after integration of Eq.~(\ref{eq:Helfrich}) over the part III as
\begin{equation}
  \label{eq:FbIII}
  F_{bIII} = \frac{1}{2} k_b \pi d  \left({{\pi\,C_{0}^2\,d}\over{2}}-{{C_{0}^2\,d}\over{2}}+
 {{16\,\pi}\over{3^{{{3}\over{2}}}\,d}}-{{8}\over{d}}+2\,\pi\,C_{0}-4
 \,C_{0}\right)
\end{equation}
The total energy can be expressed as a sum of Eq.~(\ref{eq:adhesion_energy}) and
Eqs.~(\ref{eq:FbI})--~(\ref{eq:FbIII}).

\end{document}